\documentclass[reprint,twocolumn,aps,prc,a4paper,superscriptaddress,showpacs,preprintnumbers,amsmath,amssymb]{revtex4}
\usepackage{graphicx}
\usepackage{epstopdf}
\usepackage{dcolumn}
\usepackage{mathptmx, courier, pifont}
\usepackage[scaled=0.92]{helvet}
\usepackage[T1]{fontenc}
\usepackage{textcomp}
\usepackage{color}

\begin{document}
\title{Charge and frequency resolved isochronous mass spectrometry in storage rings: First direct mass measurement of the short-lived neutron-deficient $^{51}$Co nuclide}

\author{P.~Shuai}
\affiliation{Research Center for Hadron Physics, National Laboratory of Heavy Ion Accelerator Facility in Lanzhou and University of Science and Technology of China, Hefei 230026, China}
\affiliation{Key Laboratory of High Precision Nuclear Spectroscopy and Center for Nuclear Matter Science, Institute of Modern Physics, Chinese Academy of Sciences, Lanzhou 730000, People's Republic of China}

\author{H.~S.~Xu}\thanks{Corresponding author. Email address: hushan@impcas.ac.cn}
\affiliation{Key Laboratory of High Precision Nuclear Spectroscopy and Center for Nuclear Matter Science, Institute of Modern Physics, Chinese Academy of Sciences, Lanzhou 730000, People's Republic of China}

\author{X.~L.~Tu}
\affiliation{Key Laboratory of High Precision Nuclear Spectroscopy and Center for Nuclear Matter Science, Institute of Modern Physics, Chinese Academy of Sciences, Lanzhou 730000, People's Republic of China}
\affiliation{GSI Helmholtzzentrum f\"{u}r Schwerionenforschung, Planckstra{\ss}e 1, 64291 Darmstadt, Germany}
\affiliation{Max-Planck-Institut f\"{u}r Kernphysik, Saupfercheckweg 1, 69117 Heidelberg, Germany}

\author{Y.~H.~Zhang}
\affiliation{Key Laboratory of High Precision Nuclear Spectroscopy and Center for Nuclear Matter Science, Institute of Modern Physics, Chinese Academy of Sciences, Lanzhou 730000, People's Republic of China}

\author{B.~H.~Sun}
\affiliation{School of Physics and Nuclear Energy Engineering, Beihang University, Beijing 100191, People's Republic of China}

\author{Yu.~A. Litvinov}
\affiliation{Key Laboratory of High Precision Nuclear Spectroscopy and Center for Nuclear Matter Science, Institute of Modern Physics, Chinese Academy of Sciences, Lanzhou 730000, People's Republic of China}
\affiliation{GSI Helmholtzzentrum f\"{u}r Schwerionenforschung, Planckstra{\ss}e 1, 64291 Darmstadt, Germany}
\affiliation{Max-Planck-Institut f\"{u}r Kernphysik, Saupfercheckweg 1, 69117 Heidelberg, Germany}

\author{X.~L.~Yan}
\affiliation{Key Laboratory of High Precision Nuclear Spectroscopy and Center for Nuclear Matter Science, Institute of Modern Physics, Chinese Academy of Sciences, Lanzhou 730000, People's Republic of China}
\affiliation{Graduate University of Chinese Academy of Sciences, Beijing, 100049, People's Republic of China}
\affiliation{Max-Planck-Institut f\"{u}r Kernphysik, Saupfercheckweg 1, 69117 Heidelberg, Germany}

\author{K.~Blaum}
\affiliation{Max-Planck-Institut f\"{u}r Kernphysik, Saupfercheckweg 1, 69117 Heidelberg, Germany}

\author{M.~Wang}
\affiliation{Key Laboratory of High Precision Nuclear Spectroscopy and Center for Nuclear Matter Science, Institute of Modern Physics, Chinese Academy of Sciences, Lanzhou 730000, People's Republic of China}
\affiliation{Max-Planck-Institut f\"{u}r Kernphysik, Saupfercheckweg 1, 69117 Heidelberg, Germany} \affiliation{CSNSM-IN2P3-CNRS, Universit\'{e} de Paris Sud, F-91405 Orsay, France}

\author{X.~H.~Zhou}
\affiliation{Key Laboratory of High Precision Nuclear Spectroscopy and Center for Nuclear Matter Science, Institute of Modern Physics, Chinese Academy of Sciences, Lanzhou 730000, People's Republic of China}

\author{J.~J.~He}
\affiliation{Key Laboratory of High Precision Nuclear Spectroscopy and Center for Nuclear Matter Science, Institute of Modern Physics, Chinese Academy of Sciences, Lanzhou 730000, People's Republic of China}

\author{Y.~Sun}
\affiliation{Department of Physics and Astronomy, Shanghai Jiao Tong University, Shanghai 200240, People's Republic of China}
\affiliation{Key Laboratory of High Precision Nuclear Spectroscopy and Center for Nuclear Matter Science, Institute of Modern Physics, Chinese Academy of Sciences, Lanzhou 730000, People's Republic of China}

\author{K.~Kaneko}
\affiliation{Department of Physics, Kyushu Sangyo University, Fukuoka 813-8503, Japan}

\author{Y.~J.~Yuan}
\affiliation{Key Laboratory of High Precision Nuclear Spectroscopy and Center for Nuclear Matter Science, Institute of Modern Physics, Chinese Academy of Sciences, Lanzhou 730000, People's Republic of China}

\author{J.~W.~Xia}
\affiliation{Key Laboratory of High Precision Nuclear Spectroscopy and Center for Nuclear Matter Science, Institute of Modern Physics, Chinese Academy of Sciences, Lanzhou 730000, People's Republic of China}

\author{J.~C.~Yang}
\affiliation{Key Laboratory of High Precision Nuclear Spectroscopy and Center for Nuclear Matter Science, Institute of Modern Physics, Chinese Academy of Sciences, Lanzhou 730000, People's Republic of China}

\author{G.~Audi}
\affiliation{CSNSM-IN2P3-CNRS, Universit\'{e} de Paris Sud, F-91405 Orsay, France}

\author{X.~C.~Chen}
\affiliation{Key Laboratory of High Precision Nuclear Spectroscopy and Center for Nuclear Matter Science, Institute of Modern Physics, Chinese Academy of Sciences, Lanzhou 730000, People's Republic of China}
\affiliation{Graduate University of Chinese Academy of Sciences, Beijing, 100049, People's Republic of China}

\author{G.~B.~Jia}
\affiliation{Key Laboratory of High Precision Nuclear Spectroscopy and Center for Nuclear Matter Science, Institute of Modern Physics, Chinese Academy of Sciences, Lanzhou 730000, People's Republic of China}
\affiliation{Graduate University of Chinese Academy of Sciences, Beijing, 100049, People's Republic of China}

\author{Z.~G.~Hu}
\affiliation{Key Laboratory of High Precision Nuclear Spectroscopy and Center for Nuclear Matter Science, Institute of Modern Physics, Chinese Academy of Sciences, Lanzhou 730000, People's Republic of China}

\author{X.~W.~Ma}
\affiliation{Key Laboratory of High Precision Nuclear Spectroscopy and Center for Nuclear Matter Science, Institute of Modern Physics, Chinese Academy of Sciences, Lanzhou 730000, People's Republic of China}

\author{R.~S.~Mao}
\affiliation{Key Laboratory of High Precision Nuclear Spectroscopy and Center for Nuclear Matter Science, Institute of Modern Physics, Chinese Academy of Sciences, Lanzhou 730000, People's Republic of China}

\author{B.~Mei}
\affiliation{Key Laboratory of High Precision Nuclear Spectroscopy and Center for Nuclear Matter Science, Institute of Modern Physics, Chinese Academy of Sciences, Lanzhou 730000, People's Republic of China}

\author{Z.~Y.~Sun}
\affiliation{Key Laboratory of High Precision Nuclear Spectroscopy and Center for Nuclear Matter Science, Institute of Modern Physics, Chinese Academy of Sciences, Lanzhou 730000, People's Republic of China}

\author{S.~T.~Wang}
\affiliation{Key Laboratory of High Precision Nuclear Spectroscopy and Center for Nuclear Matter Science, Institute of Modern Physics, Chinese Academy of Sciences, Lanzhou 730000, People's Republic of China}
\affiliation{Graduate University of Chinese Academy of Sciences, Beijing, 100049, People's Republic of China}

\author{G.~Q.~Xiao}
\affiliation{Key Laboratory of High Precision Nuclear Spectroscopy and Center for Nuclear Matter Science, Institute of Modern Physics, Chinese Academy of Sciences, Lanzhou 730000, People's Republic of China}

\author{X.~Xu}
\affiliation{Key Laboratory of High Precision Nuclear Spectroscopy and Center for Nuclear Matter Science, Institute of Modern Physics, Chinese Academy of Sciences, Lanzhou 730000, People's Republic of China}
\affiliation{Graduate University of Chinese Academy of Sciences, Beijing, 100049, People's Republic of China}

\author{T.~Yamaguchi}
\affiliation{Department of Physics, Saitama University, Saitama 338-8570, Japan}

\author{Y.~Yamaguchi}
\affiliation{RIKEN Nishina Center, RIKEN, Saitama 351-0198, Japan}

\author{Y.~D.~Zang}
\affiliation{Key Laboratory of High Precision Nuclear Spectroscopy and Center for Nuclear Matter Science, Institute of Modern Physics, Chinese Academy of Sciences, Lanzhou 730000, People's Republic of China}
\affiliation{Graduate University of Chinese Academy of Sciences, Beijing, 100049, People's Republic of China}

\author{H.~W.~Zhao}
\affiliation{Key Laboratory of High Precision Nuclear Spectroscopy and Center for Nuclear Matter Science, Institute of Modern Physics, Chinese Academy of Sciences, Lanzhou 730000, People's Republic of China}

\author{T.~C.~Zhao}
\affiliation{Key Laboratory of High Precision Nuclear Spectroscopy and Center for Nuclear Matter Science, Institute of Modern Physics, Chinese Academy of Sciences, Lanzhou 730000, People's Republic of China}

\author{W.~Zhang}
\affiliation{Key Laboratory of High Precision Nuclear Spectroscopy and Center for Nuclear Matter Science, Institute of Modern Physics, Chinese Academy of Sciences, Lanzhou 730000, People's Republic of China}
\affiliation{Graduate University of Chinese Academy of Sciences, Beijing, 100049, People's Republic of China}

\author{W.~L.~Zhan}
\affiliation{Key Laboratory of High Precision Nuclear Spectroscopy and Center for Nuclear Matter Science, Institute of Modern Physics, Chinese Academy of Sciences, Lanzhou 730000, People's Republic of China}


\pacs{21.10.Dr, 27.40.+z, 29.20.db}

\begin{abstract}
Revolution frequency measurements of individual ions in storage rings require sophisticated timing detectors.
One of common approaches for such detectors is the detection of secondary electrons released from a thin foil due to penetration of the stored ions.
A new method based on the analysis of intensities of secondary electrons was developed
which enables determination of the charge of each ion simultaneously with the measurement of its revolution frequency.
Although the mass-over-charge ratios of $^{51}$Co$^{27+}$ and $^{34}$Ar$^{18+}$ ions are almost identical, and therefore, the ions can not be resolved in a storage ring,
by applying the new method the mass excess of the short-lived $^{51}$Co is determined for the first time to be ME($^{51}$Co$)=-27342(48)$ keV.
Shell-model calculations in the $fp$-shell nuclei compared to the new data indicate the need to include isospin-nonconserving forces.
\end{abstract}

\maketitle




Ion storage rings are versatile facilities ideally suited for performing precision experiments with rare ions~\cite{EMIS}.
One prominent example of such experiments is the measurement of masses of short-lived nuclides~\cite{BLT}.
The masses of exotic nuclei are indispensable quantities for nuclear structure and astrophysics as well as for investigations of fundamental symmetries~\cite{Blaum1,Lunney,Grawe07,Blaum13}.
Huge efforts are undertaken worldwide, both in theory and experiment, to extend the knowledge of nuclear masses to the outskirts of the chart of nuclides.
However, since the nuclides with presently unknown masses are as a rule short-lived and/or have tiny production rates,
their measurements require very sensitive and fast techniques~\cite{Blaum2,Tu11}.
One of such techniques is the isochronous mass spectrometry (IMS) applied to nuclei of interest stored in a storage ring~\cite{FGM}.

The property of the IMS, which was pioneered at the experimental storage ring (ESR) facility of GSI \cite{Haus00,Hausmann1,Stadlmann1,Sun1}, is that the revolution
time difference $\Delta t=t-t_R$ of a stored ion with respect to a reference time $t_R$ is directly related to its mass-over-charge ratio difference $\Delta (m/q) $ by the following
expression~\cite{FGM,Haus00}:
\begin{equation}
\frac{\Delta{t}}{t_R}
=\frac{1}{\gamma_t^2}\frac{\Delta(m/q)}{(m/q)_R}, \label{eq1}
\end{equation}
where $\gamma_t$ denotes the transition point of the storage ring.
From Eq.~(\ref{eq1}) it follows that for a mass measurement it is required to achieve accurate determination of revolution times (frequencies) of stored ions.
Such determinations require sophisticated detection setups.
The revolution frequencies of stored relativistic ions are typically in the MHz range, which translates into MHz count rate on the detector.
Furthermore, since the storage rings are operated under ultra-high-vacuum conditions of $10^{-10}-10^{-12}$~mbar, the detectors have to be bake-able.
Detectors employing secondary electrons released from a thin foil by the penetrating ions are routinely employed in IMS experiments~\cite{esrdet,Mei10}.

In this Letter we present a method which allows for, in addition to the precise measurements of revolution times, a simultaneous determination of the nuclear charge of the stored ions.
Taken into account the high count rates (several MHz) and high timing accuracy (about 50~ps),
this method may be of a broad interest not only for storage-ring-based nuclear and atomic physics experiments, but also
for beam diagnostics, particle identification, and detectors for space missions, etc.
Furthermore, we present the first mass measurement of the short-lived $^{51}$Co nuclide ($T_{1/2}=68.8(1.9)$~ms \cite{dossat07}).
Since the $^{51}$Co$^{27+}$ and $^{34}$Ar$^{18+}$ ions have a very close mass-over-charge ratio ($\Delta (m/q) / (m/q) \approx 5\cdot 10^{-6}$), they cannot be resolved by their revolution times (see Eq.~(\ref{eq1})).
However, due to different atomic numbers of Co and Ar ions, the mass of $^{51}$Co could accurately be determined by applying the new method.

The experiment was performed at the HIRFL-CSR accelerator complex in the Institute of Modern Physics (IMP), Chinese Academy of Sciences (CAS),
with the combination of a heavy ion synchrotron CSRm, an in-flight fragment separator RIBLL2 and the experimental storage ring CSRe~\cite{Zhan10}.
The aim was to measure the masses of short-lived $T_z=-3/2$ nuclei in the $fp$ shell.
For this purpose, the $^{58}$Ni$^{19+}$ primary beams at an energy of 463.36~MeV/u provided by the CSRm
were fragmented in a $\sim$15~mm beryllium target placed at the entrance of RIBLL2.
The $T_z=-3/2$ fragments centred around $^{47}$Mn were transmitted through the RIBLL2 and injected into the CSRe.
The RIBLL2 and CSRe were set to a fixed magnetic rigidity of $B\rho=5.6770$~Tm.
All fragments within a B$\rho$-acceptance of $\sim\pm$0.2\% were injected into the CSRe.
The beam energy was chosen such that the fragments of interest
have the energy after the production target corresponding to $\gamma=\gamma_t =1.395$, namely the isochronous condition~\cite{Tu11,Haus00}.
At this energy the fragments emerged from the target predominantly as bare ions.

\begin{figure}[b]
\includegraphics[angle=0,width=8.5cm]{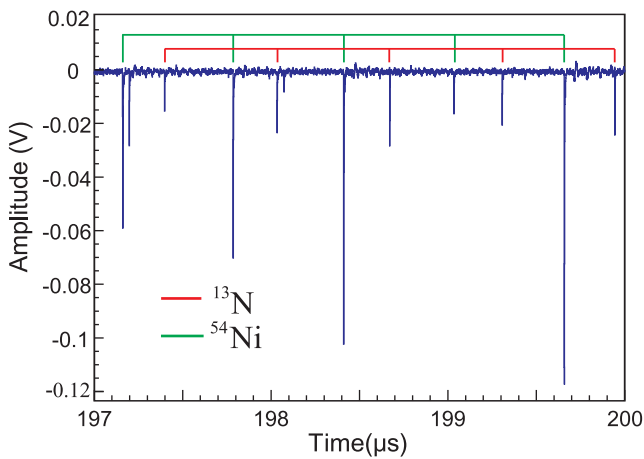}
\caption{(Colour online) Timing signals, as recorded by the time-of-flight detector, of simultaneously stored $^{13}$N$^{7+}$ and $^{54}$Ni$^{28+}$ single ions.
Only five consecutive revolutions in the CSRe are shown.
\label{Fig01}}
\end{figure}

Typically, about ten ions were stored simultaneously in one injection into the CSRe.
The revolution frequencies of stored ions were about 1.6~MHz.
In order to measure the revolution times, a dedicated timing detector
equipped with a 19~$\mu$g/cm$^2$ carbon foil of 40~mm in diameter was installed inside the CSRe.
The detector design and its operation were described in detail in Ref.~\cite{Mei10}.
When a stored ion passed through the timing detector, secondary electrons were released from the foil.
The electrons were guided isochronously by perpendicularly arranged electrostatic and magnetic fields to a set of micro-channel plates (MCP).
The signals from the MCP detector were directly put into a fast digital oscilloscope, sampled and stored for off-line analysis.
The time resolution was about 50~ps, and the detection efficiency varied from $20\%$ to $70 \%$
depending on the type and the number of stored ions~\cite{Mei10}.

\begin{figure}
\includegraphics[angle=0,width=7.5cm]{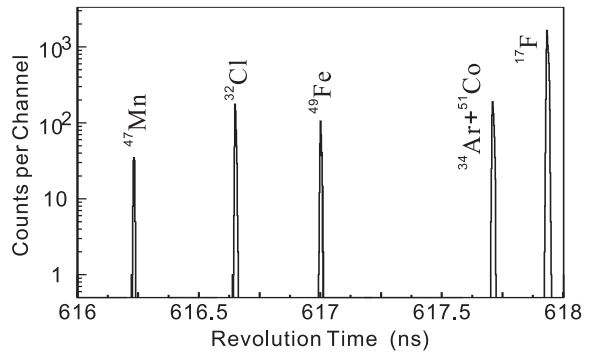}
\caption{Part of the revolution time spectrum zoomed in a time window of $616~{\rm ns} \le t \le 618$~ns.
Note that revolution time peaks of $^{51}$Co$^{27+}$ and $^{34}$Ar$^{18+}$ ions can not be resolved in the spectrum.
\label{Fig02}}
\end{figure}
The recording time was set to 200~$\mu$s for each injection which corresponds to $\approx$320 revolutions of the ions in the CSRe.
A sequence of timing signals obtained for simultaneously stored $^{13}$N$^{7+}$ and $^{54}$Ni$^{28+}$ single ions is illustrated in Figure~\ref{Fig01}.
Five revolutions corresponding to the recording time of 3~$\mu$s are depicted.

The timing signals for each individual ion are periodic.
This property was used to determine the revolution times $t_i$ of each stored ion (see Ref.~\cite{Tu11} for more details).
Only those particles which were stored for more than 300 revolutions in the CSRe were used in the analysis.
This condition is important for the precise determination of the corresponding revolution times.
All obtained revolution times were put into a histogram thus forming a revolution time spectrum.
Figure~\ref{Fig02} illustrates a part of the revolution time spectrum zoomed in a narrow time window of $616~{\rm ns} \le t \le 618$~ns.
The temporal instabilities of the magnetic fields discussed in Refs.~\cite{Tu11,TuPRL,Zhang12} have been corrected for.

Good isochronous conditions are fulfilled in a limited range of revolution times~\cite{Geissel173,Xu}.
In this experiment we analysed the range lying within $608~{\rm ns}\le t \le 620$~ns~\cite{Zhang12}, which is characterised by the
standard deviations of 2-5~ps and the optimal mass resolving power of $\Delta m/m \approx180000$ for the revolution-time peaks.
The revolution time spectrum zoomed in $608~{\rm ns}\le t \le 620$~ns is illustrated in Figure~1 of Ref.~\cite{Zhang12}.

The revolution time spectrum was calibrated by using the nuclides with accurately known masses present simultaneously in the same spectrum.
Fourteen such nuclides were used to fit $m/q$ versus $t$ employing a third order polynomial function:
\begin{equation}
\frac{m}{q}(t)=a_0+a_1\cdot{t}+a_2\cdot{t^2}+a_3\cdot{t^3}
\label{eq2}
\end{equation}
where $a_0$, $a_1$, $a_2$ and, $a_3$ are free parameters.
The masses determined in this experiment were included into the latest Atomic Mass Evaluation AME'12~\cite{AME2012}.

The peaks of $^{51}$Co$^{27+}$ and $^{34}$Ar$^{18+}$ ions in Figure~\ref{Fig02} can not be resolved in the spectrum due to very similar mass-over-charge ratios.
The standard deviations of the revolution time peaks in this range lie within 2$\sim$4~ps.
>From the systematic trends of production yields for $T_z=-3/2$ nuclides,
one would expect a significant amount of short-lived $^{51}$Co ions to be injected and stored in the CSRe.

As seen in Figure~\ref{Fig01}, signal amplitudes for ions with a larger atomic number ($^{54}$Ni$^{28+}$ ions) are on average larger than
the signals for ions with a smaller atomic number ($^{13}$N$^{7+}$ ions).
The emission of secondary electrons is well studied, see, e.g., Refs.~\cite{sel1,sel2}.
The difference in the amplitudes can be understood as a difference of ion energy loss in the detector foil, which scales in first order with $q^2$~\cite{Mei10}.

\begin{figure}
\includegraphics[angle=0,width=8.5 cm]{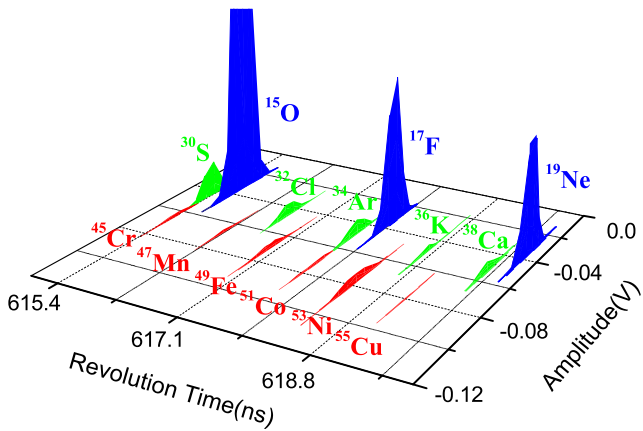}
\caption{(Colour online) Three dimensional plot (average signal amplitude vs revolution time and vs the number of stored ions)
of $^{51}$Co$^{27+}$, $^{34}$Ar$^{18+}$ and neighbouring ions.
Note that $^{51}$Co$^{27+}$ and $^{34}$Ar$^{18+}$ can not be resolved by their revolution times (see Figure~\ref{Fig02}) but have been nicely resolved by different signal amplitudes.
\label{Fig03}}
\end{figure}
To determine the mass of $^{51}$Co, we employed this $q$-dependence to resolve the peaks of $^{51}$Co$^{27+}$ and $^{34}$Ar$^{18+}$ ions.
For this purpose we analyzed the average signal amplitudes for each ion obtained from all its revolutions in the CSRe.
A two-dimensional spectrum has been produced, where the three axis are the
averaged signal height (amplitude), the revolution time, and the number of ions.
The spectrum of $^{51}$Co$^{27+}$, $^{34}$Ar$^{18+}$ and neighbouring ions is depicted in Figure~\ref{Fig03}.
One can see in this figure that the stored ions with $T_z=-1/2$, $-1$, and $-3/2$, including the $^{51}$Co$^{27+}$ and $^{34}$Ar$^{18+}$ couple,
can clearly be resolved by their signal amplitudes.
The spectrum in Figure~\ref{Fig03} is a storage ring analog of the
well-known identification plots obtained at in-flight separator (see, e.g., Figure~1 in Ref.~\cite{Hinke}).

\begin{figure}
\includegraphics[angle=0,width=8.5cm]{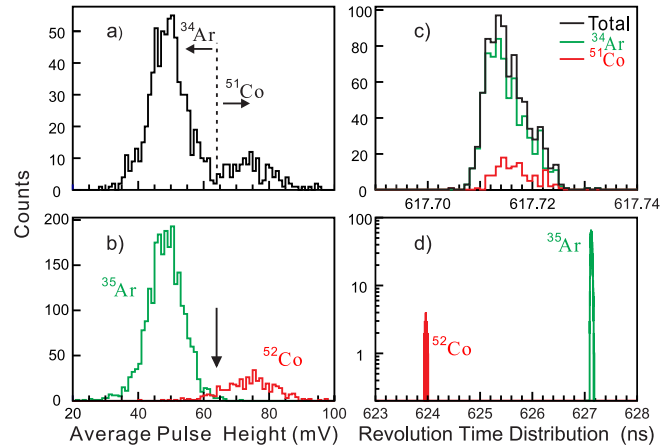}
\caption{(Color online) (a) averaged signal height distributions of
$^{34}$Ar$^{18+}$ and $^{51}$Co$^{27+}$ ions which have very close revolution times (see (c)).
(b) same as (a) but for $^{35}$Ar$^{18+}$ and $^{52}$Co$^{27+}$ ions which have
quite different revolution times (see (d)).
(c) revolution time distributions for mixed $^{34}$Ar$^{18+}$ and $^{51}$Co$^{27+}$ ions (black),
and for the individual $^{34}$Ar$^{18+}$ (green) and $^{51}$Co$^{27+}$ (red) ions obtained by
introducing gates on the signal amplitudes (see (a)).
(d) revolution time spectrum of $^{35}$Ar$^{18+}$ and $^{52}$Co$^{27+}$ ions, the signal height distributions shown in (b) are
obtained individually by setting gates on the corresponding revolution times.
\label{Fig04}}
\end{figure}
The signal height distributions for $^{51}$Co$^{27+}$ and $^{34}$Ar$^{18+}$ ions are presented in Figure~\ref{Fig04}(a).
We assume that the events with average signal amplitudes larger than 64~mV belong to $^{51}$Co$^{27+}$ ions,
while those peaks with amplitudes smaller than 64~mV are assigned to $^{34}$Ar$^{18+}$ ions.
Then, by setting these gates on each ion, the mean revolution times for $^{51}$Co$^{27+}$ and $^{34}$Ar$^{18+}$ ions
could be extracted as shown in Figure~\ref{Fig04}(c).
The corresponding revolution time difference was found to be merely 1.63(35)~ps.

Unfortunately, the peaks of $^{51}$Co$^{27+}$ and $^{34}$Ar$^{18+}$ ions
could not be completely separated from each other in the signal height distributions
(see Figs.~\ref{Fig03} and \ref{Fig04}(a)).
The remaining overlap can introduce an error in the extraction of the revolution times
which in turn can lead to an additional systematic error in the mass determination.
To estimate this influence, we investigated the signal height distributions for $^{52}$Co$^{27+}$ and $^{35}$Ar$^{18+}$ ions.
The revolution times of these ions are very different as can be seen in Figure~\ref{Fig04}(d) and
an unambiguous distribution of the averaged signal amplitudes can be extracted for the two ion species
by gating on the corresponding revolution time peak.
The latter is shown in Figure~\ref{Fig04}(b).
We emphasize that these two ion species were injected and stored simultaneously with $^{51}$Co$^{27+}$ and $^{34}$Ar$^{18+}$ ions.
By studying the obtained signal hight distribution,
we find that 10 (43) out of the 2402 (440) $^{35}$Ar$^{18+}$ ($^{52}$Co$^{27+}$) ions have
an averaged signal height larger (smaller) than 64~mV.
Since the  signal amplitude depends on $q$ of the ions, it is reasonable to assume that
distribution for the $^{35}$Ar$^{18+}$ / $^{52}$Co$^{27+}$ couple shall be the same as for the $^{34}$Ar$^{18+}$ / $^{51}$Co$^{27+}$ couple.
Hence, we can estimate that 0.4\% (10\%) of the total number of $^{34}$Ar$^{18+}$ ($^{51}$Co$^{27+}$) events
may erroneously be assigned to $^{51}$Co$^{27+}$ ($^{34}$Ar$^{18+}$).
On the basis of this estimation, we concluded that in the 704 (144) $^{34}$Ar$^{18+}$ ($^{51}$Co$^{27+}$) events,
which were used to calculate the mean revolution time of $^{34}$Ar$^{18+}$ ($^{51}$Co$^{27+}$) ions,
there could present 15(3) contaminating counts from $^{51}$Co$^{27+}$ ($^{34}$Ar$^{18+}$) corresponding to $\sim 2\%$ of the total statistics for $^{51}$Co$^{27+}$ (144 events) and $^{34}$Ar$^{18+}$ (704 events).
This mis-assignments can cause an average shift of $0.03\sim0.04$~ps in the mean revolution times,
which is negligible compared to the statistical error of the obtained time difference of 1.63(35)~ps.

The masses of $^{51}$Co and $^{34}$Ar nuclides were deduced
by interpolating the calibration curve defined by Eq.~(2) to the corresponding revolution time $t$($^{51}$Co$^{27+}$) and $t$($^{34}$Ar$^{18+}$).
The mass-excess values $ME(^{51}{\rm Co})=-27342(48)$~keV and $ME(^{34}{\rm Ar})=-18379(15)$~keV were obtained.
We note that our re-determined mass excess for $^{34}$Ar nuclide
is in excellent agreement with the independently measured value of ME($^{34}$Ar$)=-18377.2(4)$ keV~\cite{Herfurth},
providing an additional evidence for the reliability of our measured mass for $^{51}$Co.
We note, that our preliminary $ME(^{51}{\rm Co})=-27340(50)$~keV value was included into the AME'12~\cite{AME2012}.

Our accurate mass value for $^{51}$Co nuclide enables us
to determine for the first time the experimental one-proton separation energy $S_p(^{51}{\rm Co})=+142(77)$~keV.
Taking into account systematic behaviour of $S_p$ values vs mass number $A$ in this region,
$^{51}$Co is most probably the lightest proton-bound cobalt isotope with even neutron number.

\begin{figure}
\begin{center}
\includegraphics[width=0.45\textwidth]{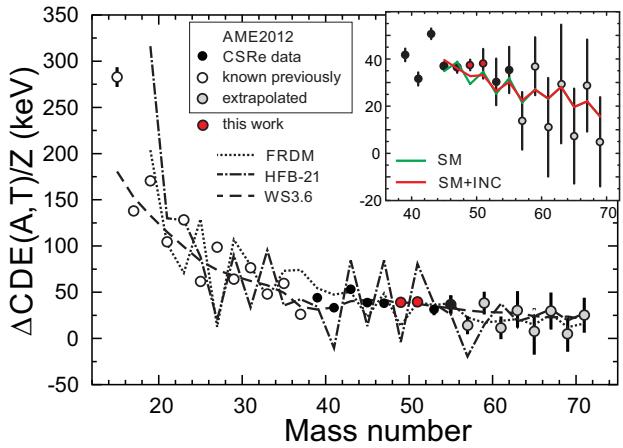}
\caption{(Colour online)
Experimental and theoretical $\Delta$CDE values.
The insert shows a zoom on the heavier mass region together
with shell model calculations without (SM) and with (SM+INC) isospin-nonconserving nuclear forces.
\label{Fig05}}
\end{center}
\end{figure}

The masses of neutron-deficient nuclides are an indispensable input for modelling the explosive nucleosynthesis processes in X-ray bursts~\cite{Parikh09,Yan13}.
However, masses of many involved nuclides are yet unknown and have to be obtained from theory.
Accurate description of nuclear masses requires correct
understanding of nuclear structure. Motivated by the observed
breakdown of the Isobaric-Multiplet Mass Equation in the
$fp$-shell~\cite{Zhang12}, we used the new mass of
$^{51}$Co to further investigate isospin-breaking
effects in the $fp$-shell.

The mass difference between mirror nuclei with isospin $T$, known as
Coulomb displacement energy (CDE), can be written as:
\begin{equation}
  CDE(A, T) = M(A,T_{Z>}) - M(A, T_{Z<}) + 2T_{Z<}\cdot\Delta_{nH} \;
  \label{eq3}
\end{equation}
where $M({A,T_{Z>}})$ and $M({A, T_{Z<}})$ are the masses of mirror
nuclei with the larger and the smaller charge, respectively.
$\Delta_{nH}=782.3466(5)$ keV is the neutron-hydrogen mass
difference~\cite{AME2012}. Extracted differential $\Delta CDE$
values, $\Delta CDE(A,T) = CDE(A+2,T)-CDE(A,T)$, for $|T_z|=3/2$ in
the $sd$- and lower $fp$-shell nuclei are plotted in
Figure~\ref{Fig05} together with predictions by often-used mass models.
Microscopic Hartree-Fock-Bogoliubov with BSk21 interaction (HFB-21)~\cite{HFB21} model as well as
macroscopic-microscopic WS-3.6 (WS)~\cite{WS3.6} and Finite-Range Droplet Model (FRDM)~\cite{Moller95} were considered.
We note that the WS model, whose $rms$ deviation from 2149 known nuclear masses is merely 336~keV,
reproduces very well the experimental data for the whole chain of Co isotopes.
The success of this model has been attributed to a better consideration of
residual correlations and isospin and mass dependence of the model parameters~\cite{WS3.6,Wang2010a,Wang2010b}.

One clearly sees in Figure~\ref{Fig05} that the
staggering of the $\Delta$CDE values, superimposed on a
monotonic decreasing curve, persists up to $A=43$,
but the effect is washed out for the heavier ones $(A=45-51)$.
The $^{51}$Co mass value is decisive here since it contributes two new data points. It is striking that
all considered mass models fail to describe the experimental staggering of
the $\Delta$CDE values. The staggering is basically not present in the
WS3.6 calculations and the HFB-21 and FRDM produce even the opposite staggering phases.

Kaneko {\it et al.}, in a study of $T=1/2$ CDE and $T=1$ TDE (triplet displacement energy) \cite{Kaneko}, pointed out
the necessary of the isospin-nonconserving (INC) nuclear interactions for the
$f_{7/2}$-shell. Now we study CDE for the $T=3/2$ isospin chain by
performing similar shell-model (SM) calculations as in Ref.~\cite{Kaneko}.
As seen in the insert of Fig.~\ref{Fig05}, the results without the INC forces indicate some staggering of $\Delta CDE$ values for $A=45-51$
while the experimental data do not show any significant staggering. By
including the INC-forces, the description of the experimental data
is improved considerably. Although not unambiguous, these results
point to the presence of INC-forces in the $pf$-shell. This
conclusion is consistent with the recent calculation for the $T=1/2$
chain~\cite{Kaneko}, where the staggering pattern was observed in the
$T=1/2$ isospin chain, but with a notable exception for the
$f_{7/2}$-shell nuclei with masses $A=42-52$. Moreover, both the
experimental data and the predictions show a similar drop at $A=53$,
where one nucleon occupies the $p_{3/2}$ orbital. Note, that the
current shell-model calculations include INC only for the
$f_{7/2}$-shell.
Our SM results show that the staggering pattern persists beyond $A=55$.
Whether and how $T=3/2$ CDE in the upper-$pf$ shell depends
on INC is an important question to be addressed by future
mass measurements of heavier $T=3/2$ nuclei.

In summary, we measured for the first time the mass of short-lived $^{51}$Co nucleus
by applying the isochronous mass spectrometry on $^{58}$Ni
projectile fragments in the storage ring CSRe in Lanzhou.
The stored $^{51}$Co$^{27+}$ and $^{34}$Ar$^{18+}$ ions can not be resolved by their revolution times due to very similar mass-over-charge ratios.
Therefore, we developed a new method in which we analysed the average amplitudes from the timing detector as a function of atomic number of the ions.
This enabled us to resolve $^{51}$Co$^{27+}$ and $^{34}$Ar$^{18+}$ ions by their charge and thus to extract their revolution times.
The newly developed method may be considered in future measurements of $Z=N$ nuclides.

The extracted proton separation energy indicates that $^{51}$Co is probably the lightest proton-bound even-$N$ Co isotope.
The behaviour of the staggering of differential $\Delta CDE$ values was used to investigate the importance of INC interactions in the $fp$-shell nuclei.
The new mass of $^{51}$Co is essential for this study since it belongs to the region ($A=45-51$) where shell model calculations turned out to be sensitive to the inclusion of INC forces.
The obtained results point to the necessity to include INC interactions in the calculations of $fp$-shell nuclei.


This work is supported in part by the 973 Program of China (No. 2013CB834401),
the NSFC (Grants No. 10925526, 11035007, U1232208, 10675147, 10805059, 11135005, 11075103, 10975008, 11105010, 11128510 and 11205205),
the Chinese Academy of Sciences, and BMBF grant in the framework of the Internationale Zusammenarbeit in Bildung und Forschung (Projekt-Nr. 01DO12012),
the External Cooperation Program of the Chinese Academy of Sciences (Grant No. GJHZ1305),
and the Helmholtz-CAS Joint Research Group (Group No. HCJRG-108).
Y.A.L is supported by CAS visiting professorship for senior international scientists (Grant No. 2009J2-23).
K.B. acknowledges support by the Nuclear Astrophysics Virtual Institute (NAVI) of the Helmholtz Association.
K.B. and Y.A.L. thank ESF for support within the EuroGENESIS program.
T.Y. acknowledges support by The Mitsubishi Foundation.


\begin{thebibliography} {99}
\vspace{-0.5cm}
\bibitem{EMIS}
Yu.~A.~Litvinov {\it et al.}, Nucl. Instr. Meth. B {\bf 317}, 603 (2013).

\bibitem{BLT}
F.~Bosch, Yu.A.~Litvinov, and Th. St{\"o}hlker, Prog. Part. Nucl. Phys. {\bf 73}, 84 (2013).

\bibitem{Blaum1}
K.~Blaum, Phys.~Rep.~{\bf 425}, 1 (2006).

\bibitem{Lunney}
D.~Lunney {\it et al.} Rev.~Mod.~Phys.~{\bf 75}, 1021 (2003).

\bibitem{Grawe07}
H.~Grawe, K.~Langanke, and G.~Mart\'{\i}nez-Pinedo, Rep.~Prog.~Phys.~ {\bf 70}, 1525 (2007).

\bibitem{Blaum13}
K.~Blaum, J.~Dilling, and W. N{\"o}rtesh{\"a}user, Phys. Scripta T {\bf 152}, 014017 (2013).

\bibitem{Blaum2}
K.~Blaum {\it et al.}, J.~Phys.~Conf.~Series~{\bf 312}, 092001 (2011).

\bibitem{Tu11}
X.~L.~Tu {\it et al.}, Nucl.~Instr.~Meth.~A~{\bf 654}, 213 (2011).

\bibitem{FGM}
B.~Franzke, H.~Geissel and G.~M\"{u}nzenberg, Mass Spec. Rev.~{\bf 27}, 428 (2008).

\bibitem{Haus00}
M.~Hausmann {\it et al.}, Nucl.~Instr.~Meth.~{\bf A 446}, 569 (2000).

\bibitem{Hausmann1}
M.~Hausmann, {\it et al.}, Hyperfine Interactions~{\bf 132}, 291 (2001).

\bibitem{Stadlmann1}
J.~Stadlmann, {\it et al.}, Phys.~Lett.~B~{\bf 586}, 27 (2004).

\bibitem{Sun1}
B.~Sun, {\it et al.}, Nucl.~Phys.~A~{\bf 812}, 1 (2008).

\bibitem{esrdet}
J.~Tr{\"o}tscher {\it et al.}, Nucl. Instr. Meth. B {\bf 70}, 455 (1992).

\bibitem{Mei10}
B.~Mei, {\it et al.}, Nucl.~Instr.~and Meth. A~{\bf 624}, 109 (2010).

\bibitem{dossat07}
C.~Dossat {\it et al.}, Nucl. Phys. A {\bf 792}, 18 (2007).

\bibitem{Zhan10}
W.~L.~Zhan {\it et al.}, Nucl.~Phys.~A~{\bf 834}, 694c (2010).

\bibitem{TuPRL}
X.~L.~Tu {\it et al}, Phys. Rev. Lett. {\bf 106}, 112501 (2011).

\bibitem{Zhang12}
Y.~H.~Zhang {\it et al.}, Phys. Rev. Lett. {\bf 109}, 102501 (2012).

\bibitem{Geissel173}
H.~Geissel {\it et al.}, Hyperfine Interactions {\bf 173}, 49 (2006).

\bibitem{Xu}
H.~S.~Xu {\it et al.}, Int. J. Mass Spectrom {\bf 349-350}, 162 (2013).
\bibitem{AME2012}
M.~Wang, G.~Audi, A.H.~Wapstra {\it et al.}, Chin. Phys. C {\bf 36}, 1603 (2012).

\bibitem{sel1}
A.~Clouvas {\it et al.}, Phys. Rev. B {\bf 55}, 55 (1997).

\bibitem{sel2}
H.~P.~Garnir {\it et al.}, Nucl. Instr. Meth. {\bf 202}, 187 (1982).

\bibitem{Hinke}
C.~B.~Hinke {\it et al.}, Nature {\bf 486}, 341 (2012).

\bibitem{Herfurth}
F.~Herfurth {\it et al.}, Phys. Rev. Lett.  {\bf 87}, 142501 (2001).

\bibitem{Parikh09}
A. Parikh {\it et al.}, Phys. Rev. C {\bf 79}, 045802 (2009)

\bibitem{Yan13}
X.~L.~Yan {\it et al.}, Astrop. J. Letters {\bf 766}, L8 (2013)


\bibitem{HFB21}
S. Goriely, N. Chamel, J.M.~Pearson, Phys. Rev. C. 82, 035804 (2010).

\bibitem{WS3.6}
M.~Liu, N.~Wang, Y.~Deng, X.~Wu, Phys. Rev. C 84 (2011) 014333.

\bibitem{Moller95}
P.~M\"{o}ller, {\it et al.}, At. Data Nucl. Data Tables~{\bf 59},185 (1995).

\bibitem{Wang2010a}
N. Wang, M. Liu and X. Wu, Phys. Rev. C 81, 044322 (2010).

\bibitem{Wang2010b}
N.Wang, Z. Liang, M. Liu and X.Wu, Phys. Rev. C 82, 044304 (2010).


\bibitem{Kaneko}
K.~Kaneko, Y.~Sun, T. Mizusaki, and S. Tazaki, Phys. Rev. Lett. {\bf 110}, 172505 (2013).




\end{thebibliography}
\end{document}